# Tunable Quantum Anomalous Hall Effect via Crystal Order in Spin-Splitting Antiferromagnets

*Wenxuan Zhu,[†*] Hua Bai,[†] Lei Han, Feng Pan Cheng Song[*]*

[1] Key Laboratory of Advanced Materials, School of Materials Science and Engineering, Tsinghua University, Beijing 100084, China

[†] These authors contributed equally: Wenxuan Zhu, Hua Bai

[*] Corresponding author. Email: zhu-wx19@tsinghua.org.cn; songcheng@mail.tsinghua.edu.cn

**ABSTRACT:** Quantum anomalous Hall (QAH) effect provides dissipationless chiral channels for spin transport, expected as an outstanding candidate in future low-power quantum computation. The spin-splitting band structure is vital for obtaining QAH effect in topological systems, with ferromagnetism indispensable to manipulate the Chern number. Herein, we challenge this wisdom by proposing tunable QAH effect in spin-splitting antiferromagnets with zero magnetization. Since the spin splitting of these unique magnets originates from the alternate crystal environment, the Chern number can be modulated not only by the conventional magnetic order, but also by the crystal order, opening an additional dimension for tuning QAH effect. Our concept is illustrated based on two-dimensional (2D) $MnBi_2Te_4$ (MBT) with even septuple layers (SLs), a typical axion insulator with fully magnetic compensation. By interlayer rotation and translation operations, sublattices of MBT with opposite magnetizations are no longer connected by inversion or mirror symmetries, leading to the transition to the QAH insulator. The flexible stacking of 2D materials enables the reversible Chern number by crystal design. Our work fundamentally reveals the crystal-order-dependent QAH effect in spin-splitting antiferromagnets, which would advance QAH effect-based devices towards high controllability, integration density and operation speed.

**KEYWORDS:** quantum anomalous Hall effect, spin-splitting antiferromagnet, $MnBi_2Te_4$, tunable Chern number, crystal design

Quantum anomalous Hall (QAH) effect shows chiral edge states with topological protection against local perturbations, which can be characterized by quantized Hall conductance with zero longitudinal conductance.[1,2] Since it provides dissipationless chiral channels for transport, QAH effect is highly expected to realize low-power electronic devices and future quantum computation,[3–5] which makes the achievement of QAH effect a recognized target in the fields of both condensed matter physics and materials science. In QAH insulators, the topologically nontrivial band structures with spin splitting are generally required to achieve the chiral edge states. The recognized perspective is introducing ferromagnetism to topological insulators, which breaks the global time-reversal symmetry and induces Zeeman splitting in band structures (left panel in Figure 1a). Guided by this idea, QAH effect has been experimentally observed in magnetic atoms-doped topological insulators[6–9] and moiré superlattices with magnetic ordering induced by interactions.[10–12] Especially, in two-dimensional (2D) intrinsic magnetic topological insulator $MnBi_2Te_4$ (MBT), odd septuple layers (SLs)[13] or superlattices[14] are necessary to obtain ferromagnetism and resultant QAH effect, while even-SL MBT is only axion insulator.[15] Such reliance of QAH effect on Zeeman spin splitting results in the only determination of Chern number by the ferromagnetic ordering.

Antiferromagnetic materials without net magnetization and resultant stray fields have been demonstrated to promote the integration density and robustness of devices, whose intrinsic terahertz dynamics also enable fast operation.[16,17] Therefore, the realization of QAH effect in antiferromagnets is highly expected to advance QAH effect-based devices but is severely obstructed by the accompanied spin-degenerated band structures in traditional antiferromagnets.[18,19] Fortunately, a special category of magnets that combines spin-splitting band structures and compensated antiferromagnetic ordering has recently been proposed and experimentally verified.[18–21] For these spin-splitting antiferromagnets, the lift of spin degeneracy originates from the special crystal symmetry with *C*-paired spin momentum locking, rather than the Zeeman splitting,[18,19,22] giving rise to alternate spin splitting in the reciprocal space, as schematically shown in the right panel of Figure 1a. Such origin establishes a close connection between spin splitting bands with the crystal symmetry, manifesting their properties with the dependence on the crystal order.[23,24] Therefore, spin-splitting antiferromagnets show high potential to achieve QAH effects with the advantages of antiferromagnets. More importantly, it is promising to realize the control of Chern number by crystal order besides magnetic order in spin-splitting antiferromagnets. However, the covalent bonding between the alternate crystal environment in investigated spin-splitting antiferromagnets impedes the flexible manipulation of crystal order, making such potential still unrevealed.

Herein, we theoretically propose the achievement of QAH effect in spin-splitting antiferromagnets with Chern number defined by the crystal order, fulfilled by taking the advantages of 2D layered structure in MBT. MBT possesses perpendicular magnetic anisotropy with interlayer antiferromagnetic coupling[25] due to the half-occupied 3*d* orbitals of $Mn^{2+}$. Mn atom of each SL in MBT occupies the center of the squeezed octahedral composed of Te atoms, which results in anisotropic magnetization density around Mn atoms (left panel of Figure 1b). The magnetization density along the $r_2$ direction is more significant than that along the $r_1$ direction. In pristine MBT, the anisotropy of magnetization densities in adjacent two SLs are the same due to the same orientation of octahedral crystal environment, resulting in spin degeneracy. When one of the sublattice layers is rotated 180º around the *c* axis, the octahedral crystal environment of adjacent SLs is altered, leading to the magnetization densities with opposite anisotropies (right panel of Figure 1b). Accordingly, the energy bands are expected to exhibit spin splitting in this system. The interlayer translation that breaks the mirror symmetry is furtherly introduced to lift the degeneracy at the Γ point in *k* space, where the topologically nontrivial bands exactly occur. Due to the van der Waals bonding between the alternate crystal environment, the crystal order can be flexible controlled by the direction of interlayer translation, which determines the polarity of spin splitting bands. Guided by this strategy, we construct even-SL MBT, whose energy bands simultaneously possess topologically nontrivial properties and spin splitting. Correspondingly, the achievement of QAH effect is theoretically demonstrated, whose Chern number is switchable between +1 and −1 by both the crystal order and Néel order.

The design of spin-splitting antiferromagnets with tunable spin splitting based on MBT is first demonstrated in 2 SLs. In pristine 2-SL MBT, the top layer is shifted $\frac{2}{3}\boldsymbol{a}+\frac{1}{3}\boldsymbol{b}$ relative to the bottom layer, creating the stacking order named by AB (Figure 1c). With interlayer antiferromagnetic coupling and the same octahedral crystal environment between the SLs, the spin-up and spin-down bands are degenerated (Figure S1). After rotating the top layer 180 degrees around the *c* axis to obtain different crystal fields in two SLs, three stacking orders are correspondingly created: without interlayer translation, and with the top layer shifted by $\frac{2}{3}\boldsymbol{a}+\frac{1}{3}\boldsymbol{b}$ and $\frac{1}{3}\boldsymbol{a}+\frac{2}{3}\boldsymbol{b}$ relative to the bottom layer, named as AA[R], AC[R] and AB[R], respectively. Due to the half-occupied 3*d* orbitals of $Mn^{2+}$, interlayer ferromagnetic coupling built by interlayer hopping is forbidden[26], which leads to the same ground state of antiferromagnetic ordering in all these interlayer stacking orders (Table S1). In AA$^R$ stacking order, despite of the existence of spin splitting induced by different crystal environment around Mn atoms in the two SLs (Figure S2), the mirror symmetry is preserved between two layers, making the spin splitting negligible at the Γ point in *k* space (Figure 1d). In contrast, the mirror symmetry is broken by the interlayer translation transformation in AC$^R$ and AB$^R$ stacking orders, satisfying the requirements of introducing spin splitting at the Γ point. Additionally, according to the calculations of total energy (Table S2), 2-SL MBT with AC$^R$ and AB$^R$ stacking orders show much higher stability compared with AA$^R$ stacking orders. More importantly, the total energy of both AC$^R$ and AB$^R$ stacking orders is only around 10 meV/u. c. higher than that of pristine 2-SL MBT. Considering that AC$^R$ and AB$^R$ stacking orders stand a good chance of stable existence in experiments, we mainly focus on these two stacking orders in the discussion below.

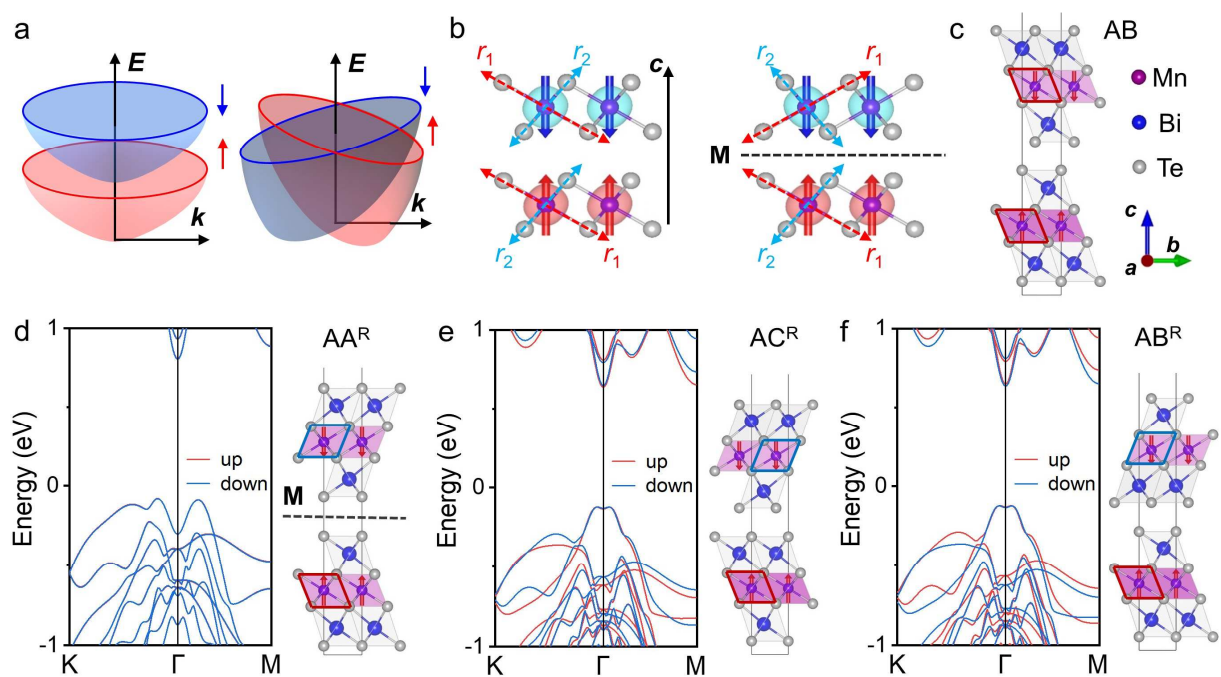

**Figure 1.** Stacking-dependent spin spitting in 2-SL MBT. (a) Schematics of spin splitting induced by ferromagnetism and alternate octahedral crystal fields. Opposite magnetizations

and spin-polarized bands are distinguished by red and blue arrows. (b) Magnetization densities in pristine MBT (left panel) and MBT with interlayer rotation transformation (right panel). The magnetization densities of opposite local moments are shown in red and blue, respectively. The interlayer mirror symmetry in the right panel is represented by the black dashed line. (c) Atomic structure of intrinsic 2SL-MBT with interlayer AB stacking order. The orientation of octahedral structures around Mn atoms are highlighted by red and blue diamonds. (d−f) Spin-dependent band structures rotated bilayer MBT with $AA^R$ (d) $AC^R$ (e) and $AB^R$ (f) stacking order. The atomic structures are shown on the left side of the band structures. Mn, Bi and Te atoms are represented by purple, blue and grey atoms, respectively. The red arrows represent the directions of local moments. The orientation of octahedral structures around Mn atoms are highlighted by red and blue diamonds.

Figure 1e shows the crystal structure and spin-dependent band structures of 2-SL MBT with $AC^R$ stacking. In this case, the spin splitting is much more significant compared with the $AA^R$ stacking order, especially at the M point in $k$ space. More importantly, although the spin splitting at the Γ point is relatively small, it is also distinguishable. The spin splitting is reversed when switching the magnetization in the two layers simultaneously, namely reversing the Néel order (Figure S3). This demonstrates that the different crystal environment between opposite magnetizations created by interlayer rotation transition effectively induces spin-splitting bands. Additionally, the interlayer translation transition breaks the mirror symmetry, bringing about the occurrence of spin splitting at the Γ point. Consequently, the request of spin-splitting bands at the Γ point to realize QAH effect is fulfilled in compensated antiferromagnets with zero net magnetization. Besides the Néel order, the spin splitting is also determined by the crystal order[23,24]. In $AB^R$ stacking order, the top layer is shifted by $\frac{1}{3}\boldsymbol{a}+\frac{2}{3}\boldsymbol{b}$ relative to the bottom layer. According to the crystal symmetry of MBT monolayer, $AB^R$ stacking order is equivalent to shifting the top layer in the opposite direction compared with $AC^R$ stacking order (Figure S4). Correspondingly, the spin splitting in band structures of 2-SL MBT with $AB^R$ stacking order is found to be opposite to that of $AC^R$ stacking order as illustrated by Figure 1f, where the splitting energy in *k* space is almost the same. The opposite spin splitting in $AC^R$ and $AB^R$ stacking orders is further supported by the spin density of states (Figure S5). Differently, in pristine MBT without interlayer rotation and resultant spin splitting, the AC and AB stacking orders are almost completely equivalent, where the interlayer translations are consistent with those of $AC^R$ and $AB^R$ stacking orders. Therefore, the creation of spin splitting at the Γ point of 2-SL MBT with compensated magnetization is achieved by interlayer rotation and translation transformation, which is reversible by changing the Néel order and crystal order.

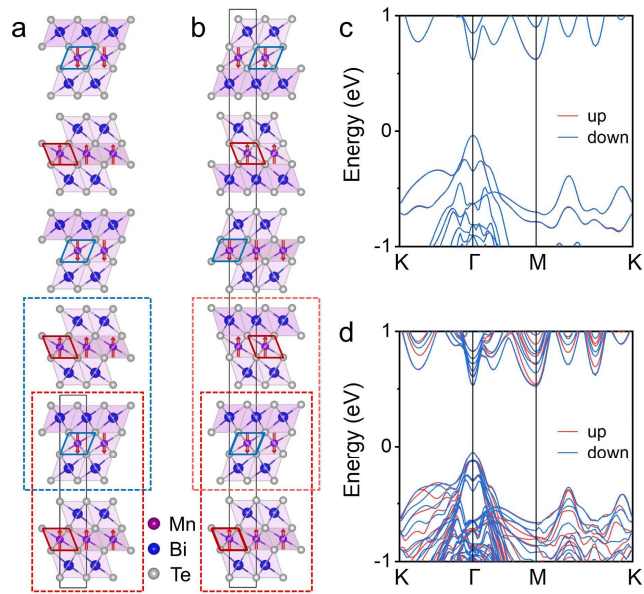

**Figure 2.** Stacking-dependent spin splitting in bulk MBT. (a,b) Atomic structure of bulk MBT constructed by the unit cells of AC$^R$ stacked 2-SL MBT, which are stacked in the order of A-A (a) and A-C-B-A (b). The primitive cells of the two bulk structures are both highlighted by black rectangles. The stacking orders in the unit cells in both bulk structures are highlighted by red dashed rectangles. The stacking orders between unit cells in (a) and (b) are highlighted by blue and light red dashed rectangles, respectively. The directions of magnetizations are represented by red arrows. (c,d) Spin-dependent band structures of bulk MBT with the structure shown in (a) and (b).

Considering that the topologically nontrivial bands are generally restricted in 2-SL MBT due to the quantum confinement effects, the structures are expanded from 2 SL to bulk and multilayers based on the design of interlayer stacking. Firstly, two structures of bulk can be constructed with the unit cell of AC$^R$ stacked 2-SL MBT, which are shown in Figure 2a and b. In the first bulk structure (Figure 2a), the unit cells are stacked with the order of A-A, where the unit cell is the primitive cell. In this case, despite the alternate crystal environment around opposite magnetizations between adjacent SLs, the spin splitting in the band structure of bulk becomes negligible as shown in Figure2c. Through the analysis of spin-dependent band structure (Figure S6), the spin splitting induced by the interlayer translation between two primitive cells (highlighted by blue in Figure 2a) is opposite to that of the AC$^R$ stacking order in the primitive cell (highlighted by red in Figure 2a). The opposite spin splitting induced by adjacent interlayers brings about spin degeneracy in bulk. Therefore, the interlayer translation also needs to be conserved to maintain the significant spin splitting in bulk or multilayers.

Consequently, the bulk structure in Figure 2b is constructed, where the unit cells are stacked in the order of A-C-B-A. In this case, the interlayer translations in the unit cell and between the unit cells are all consistent, where the top SL is shifted by $\frac{2}{3}\bm{a}+\frac{1}{3}\bm{b}$ relative the bottom SL. The interlayer transformations between each 2 SLs are the same as those of $AC^R$ stacking order, which consequently maintains the spin-splitting band structure (Figure 2d). It can also be found that this bulk structure with interlayer $AC^R$ stacking order (Figure 2b) can be equivalent to the bulk structure of pristine MBT with the even SLs rotated 180 degrees around the $c$ axis.

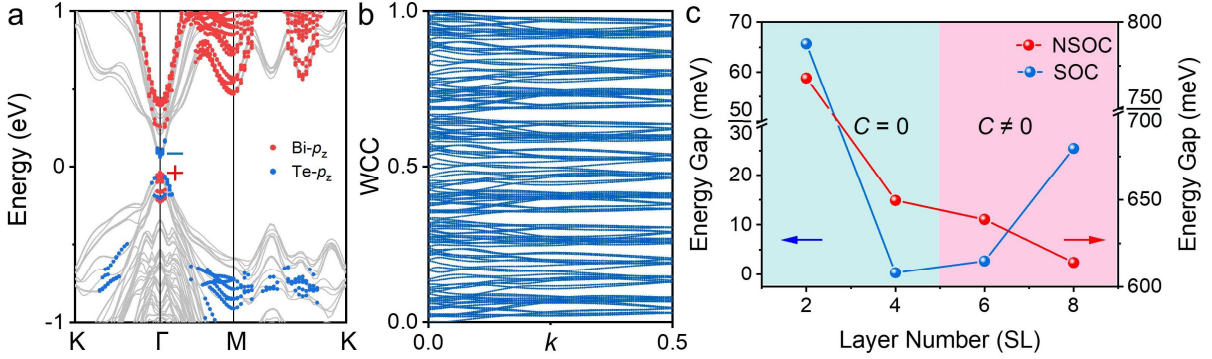

**Figure 3.** Topological properties of bulk and multilayer MBT with interlayer $AC^R$ stacking order. (a) Band structures of bulk MBT with interlayer $AC^R$ stacking order with SOC. The occupations of Bi-$p_z$ and Te-$p_z$ orbitals are highlighted by red and blue, respectively. (b) Evolution of calculated WCCs in $k_z = 0$ plane. (c) Layer number-dependent energy gap of band structures with SOC (left axis) and without SOC (NSOC, right axis), and the topological Chern number in multilayer MBT with interlayer $AC^R$ stacking order.

The topological properties of bulk and multilayers MBT with interlayer $AC^R$ stacking order and antiferromagnetic coupling shown in Figure 2b are then revealed. The topological nontrivial properties of bulk situation are first demonstrated by the band structures with orbital projections. As shown in Figure 3a, the occupations by the Bi-$p_z$ and Te-$p_z$ orbitals at the valance band maximum and conduction band minimum are inversed with the consideration of spin-orbit coupling (SOC). This reflects that the bulk MBT with interlayer AC$^R$ stacking order maintains topological nontrivial band structures [25,27]. The calculated evolution of Wannier Charge Centers (WCCs) in the plane of $k_z = 0$ indicates a nonzero topological invariant $Z_2 = 1$ with SOC[28,29], further demonstrating that this bulk system is the three-dimensional antiferromagnetic topological insulator.

With the revelation of topological nontrivial properties in bulk situation, the thickness is reduced to multilayers to obtain QAH insulators with a nonzero Chern number. In multilayers, the topological properties vary with the layer number. To exclude the spin-splitting induced from the uncompensated net magnetization, we focus on the multilayer systems with even SLs. One of the representative features of topological nontrivial properties is band inversion induced by SOC. Thus, the occurrence of band inversion depends on the SOC strength and the energy gap without SOC, while the latter one is determined by the strength of interlayer orbital hybridization. The band structures of multilayers without SOC are calculated (Figure S7) and the energy gaps are extracted. As shown in Figure 3c, the energy gap of band structures without SOC gradually decreases with the increase of the layer number due to the enhanced interlayer hybridization in multilayers. After the inclusion of SOC, the energy gap first decreases with the increase of the layer number below 4 SL and approaches zero at 4 SL. In this regime, the strength of SOC is unable to induce band inversion at the Γ point of $k$ space (Figure S8). The topological trivial band structure leads to Chern number $C = 0$ below 4 SL, further supported by the calculation of WCCs (Figure S9). When further increasing the layer number to 6 SL, the bandgap reopens with the band inversion occurs at the Γ point (Figure S8), indicating the topologically nontrivial band structures. Given that this system simultaneously exhibits the topologically nontrivial and spin-splitting bands at the Γ point, $AC^R$-stacked MBT beyond 6 SLs becomes a QAH insulator, which possesses nonzero Chern number. The Chern numbers of multilayers are further supported by the calculations of WCCs (Figure S9). The representative calculated evolution of WCCs of 8-SL MBT in $k_z = 0$ plane with a closed momentum surface is shown in Figure 4a, where the WCCs of occupied bands cover the plane with almost the same polarization of evolution. The summation of WCCs highlighted by the red line defines the total polarization of evolution and reflects the topological invariant of $C = 1$[28,30], suggesting the achievement of QAH effect.

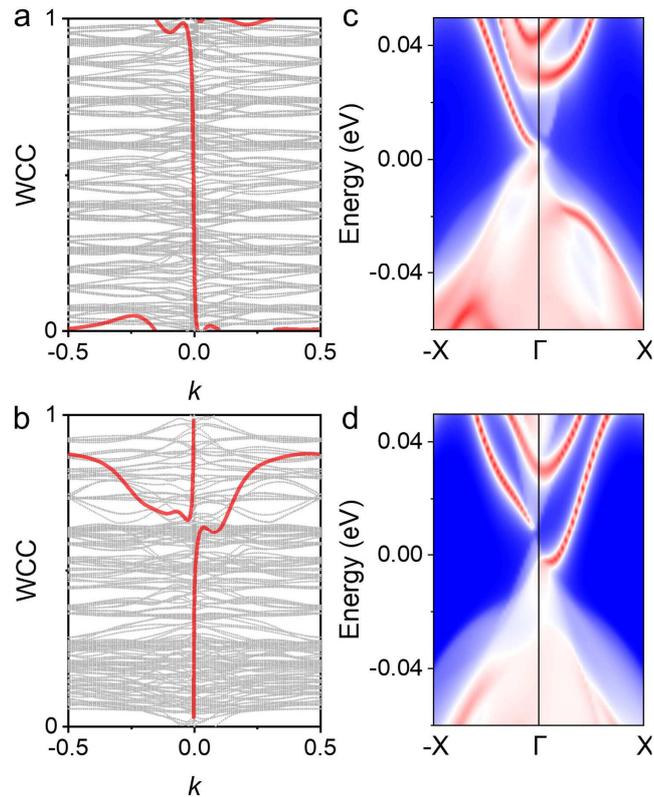

**Figure 4.** QAH state in $AC^R$ and $AB^R$ stacked 8-SL MBT. (a,b) Calculated evolutions of WCCs in $k_z = 0$ plane with a closed momentum surface of 8-SL MBT with interlayer $AC^R$ (a)

and $AB^R$ (b) stacking orders. The summation of WCCs is highlighted in red. (c,d) Edge states of 8-SL MBT with interlayer $AC^R$ (c) and $AB^R$ (d) stacking orders.

The sign of Chern number is an important feature of QAH effect, which reflects its polarization and is generally controlled by changing the direction of net magnetization in the system[6,10,13]. Here, the Chern number can be reversed by changing the Néel order (Figure S10), which can control the spin splitting. More interestingly, since the spin splitting originates from the alternate crystal environment, the crystal order also determines the spin splitting and correspondingly controls the sign of Chern number. To verify such effect, the evolution of WCCs in 8-SL with interlayer $AB^R$ stacking order is calculated. As illustrated by Figure 4b, the evolution of WCCs in $k_z = 0$ plane with a closed momentum surface also represents a nonzero Chern number, indicating the existence of QAH effect. More importantly, the polarity of evolution in Figure 4b, defined by the summation, is opposite to that shown in Figure 4a. Such feature unveils the opposite nonzero Chern number[31] in 8-SL MBT with different interlayer translation transformations, namely $C = 1$ in $AC^R$ stacking order and $C = -1$ in $AB^R$ stacking order. QAH effect with the opposite Chern number is further supported by the calculated edge states, which are shown in Figures 4c and d. The chiral edge states across the band gap can be clearly observed in the surface states of both $AC^R$ (Figure 4c) and $AB^R$ (Figure 4d) stacking orders, reflecting both systems are QAH insulators. Differently, the polarities of edge states from the same surface are opposite in the two stacking orders, demonstrating the opposite Chern number. The introduction of spin-splitting antiferromagnets opens up the dimension of crystal order for modulating the polarization of QAH effect.

Due to the inherent layered structure of 2D materials with interlayer van der Waals gaps, the variable stacking order shows significance in the determination of intrinsic properties.[32,33] Correspondingly, many methods have been developed to obtain different interlayer stacking orders in 2D materials, including hydrostatic pressure[34,35], molecular beam epitaxy[36], electric fields[37] and strain[38]. The development of these techniques supports the experimental realization of MBT with designed stacking orders, which shows only small energy difference compared to intrinsic MBT.

In conclusion, we have proposed the crystal-defined QAH effect in spin-splitting antiferromagnets with fully magnetic compensation. By taking advantage of the flexible buildability in 2D materials, interlayer rotation and translation transformations are introduced in even-SL MBT with collinear antiferromagnetic ordering. The creation of alternated crystal fields between layers and the broken mirror symmetry induces spin splitting in topological nontrivial bands. The polarity of spin splitting is flexibly tunable by the interlayer translation, which determines the crystal order. QAH effect is consequently achieved with Chern number reversible between +1 and −1 according to the crystal order. Our work manifests the uniqueness of spin splitting antiferromagnets, broadening the systems of QAH effect and the modulation dimensions of Chern number. Benefitted from the advantages of absent stray field and high intrinsic frequency in antiferromagnets, the integration density and operation speed of QAH effect-based devices will also be effectively promoted.

**Methods**

The first-principles calculations were performed based on the Vienna ab initio simulation package (VASP)[39,40] with projector augmented wave method[41,42]. The exchange correlation interaction is treated by the Perdew-Burke-Ernzerhof (PBE) functional[43] with the energy cutoff set as 350 eV. The GGA+U method was used to treat the localized 3$d$ orbitals, in which the value of $U_{eff}$ is set as 4 eV for the 3$d$ orbitals of Mn element, referring to previous studies on MBT[25,44]. DFT-D3 was used to properly treated the interlayer van der Waals interaction in all systems[45]. The structures of bulk systems are optimized with both the lattice constants and atom positions fully relaxed, using the force criterion of 0.01 eV/ Å. A vacuum layer larger than 20 Å was adopted in all calculations of thin films. The calculations of WCCs and the surface states were performed with WannierTools package based on maximally localized Wannier functions[46].

ASSOCIATED CONTENT

**Supporting Information.** Calculations of the total energy and exchange energy; Spin-dependent band structures and density of states of different stacking orders; Schematic of interlayer translation transformations; Band structures of multilayers with and without SOC; WCCs of multilayers; WCCs of 8-SL MBT with reversed Néel order.

AUTHOR INFORMATION

**Corresponding Author**

Wenxuan Zhu. Key Laboratory of Advanced Materials (MOE), School of Materials Science and Engineering, Tsinghua University, Beijing 100084, China. Email: zhu-wx19@tsinghua.org.cn

Cheng Song. Key Laboratory of Advanced Materials (MOE), School of Materials Science and Engineering, Tsinghua University, Beijing 100084, China. Email: songcheng@mail.tsinghua.edu.cn

**Author contributions**

†W. Z., H. B. contributed equally to the work. C.S. and F.P. led the project. W.Z. and H.B. proposed the study. W.Z. did the theoretical calculations. L.H. provided comments on the calculations. W.Z., H.B. wrote the manuscript. L.H., C.S. and F.P. reviewed and optimized the manuscript. All authors discussed the results and commented on the manuscript.

**Notes**

The authors have no conflicts to disclose.

ACKNOWLEDGEMENT

This work was supported by the National Natural Science Foundation of China (Grant Nos. 52225106, 12241404).